\RequirePackage{lineno}
\documentclass[aps,prl,showpacs,twocolumn,superscriptaddress,groupedaddress,floatfix,amsfonts]{revtex4}
\usepackage{graphicx}
\usepackage{dcolumn}
\usepackage{bm}
\usepackage{amssymb}
\usepackage{amsmath}
\usepackage{xspace}
\newcommand{\la}{\langle}
\newcommand{\ra}{\rangle}
\newcommand{\lt}{\!<\!}

\newcommand{\GeV}{\ensuremath{\text{GeV}}\xspace}
\newcommand{\TeV}{\ensuremath{\text{TeV}}\xspace}
\newcommand{\Ptg}{p_{T}^{\gamma}\xspace}
\newcommand{\ptg}{\ensuremath{p_T^{\gamma}}\xspace}
\newcommand{\gb}{\ensuremath{\gamma+{b}}\xspace}

\usepackage{color}
\newcommand{\CHECK}[1]{\textbf{\color{red}[#1]}\xspace}
\newcommand{\chk}[1]{\CHECK{CHECK THIS!}}

\begin{document}
\hspace{5.2in} \mbox{FERMILAB-PUB-14-135-E}

\title{\boldmath Measurement of the differential $\gamma+2~b$-jet cross section and 
the ratio $\sigma$($\gamma+2~b$-jets)/$\sigma$($\gamma+b$-jet) in $p\bar{p}$ collisions at $\sqrt{s}=1.96~\TeV$}
\affiliation{LAFEX, Centro Brasileiro de Pesquisas F\'{i}sicas, Rio de Janeiro, Brazil}
\affiliation{Universidade do Estado do Rio de Janeiro, Rio de Janeiro, Brazil}
\affiliation{Universidade Federal do ABC, Santo Andr\'e, Brazil}
\affiliation{University of Science and Technology of China, Hefei, People's Republic of China}
\affiliation{Universidad de los Andes, Bogot\'a, Colombia}
\affiliation{Charles University, Faculty of Mathematics and Physics, Center for Particle Physics, Prague, Czech Republic}
\affiliation{Czech Technical University in Prague, Prague, Czech Republic}
\affiliation{Institute of Physics, Academy of Sciences of the Czech Republic, Prague, Czech Republic}
\affiliation{Universidad San Francisco de Quito, Quito, Ecuador}
\affiliation{LPC, Universit\'e Blaise Pascal, CNRS/IN2P3, Clermont, France}
\affiliation{LPSC, Universit\'e Joseph Fourier Grenoble 1, CNRS/IN2P3, Institut National Polytechnique de Grenoble, Grenoble, France}
\affiliation{CPPM, Aix-Marseille Universit\'e, CNRS/IN2P3, Marseille, France}
\affiliation{LAL, Universit\'e Paris-Sud, CNRS/IN2P3, Orsay, France}
\affiliation{LPNHE, Universit\'es Paris VI and VII, CNRS/IN2P3, Paris, France}
\affiliation{CEA, Irfu, SPP, Saclay, France}
\affiliation{IPHC, Universit\'e de Strasbourg, CNRS/IN2P3, Strasbourg, France}
\affiliation{IPNL, Universit\'e Lyon 1, CNRS/IN2P3, Villeurbanne, France and Universit\'e de Lyon, Lyon, France}
\affiliation{III. Physikalisches Institut A, RWTH Aachen University, Aachen, Germany}
\affiliation{Physikalisches Institut, Universit\"at Freiburg, Freiburg, Germany}
\affiliation{II. Physikalisches Institut, Georg-August-Universit\"at G\"ottingen, G\"ottingen, Germany}
\affiliation{Institut f\"ur Physik, Universit\"at Mainz, Mainz, Germany}
\affiliation{Ludwig-Maximilians-Universit\"at M\"unchen, M\"unchen, Germany}
\affiliation{Panjab University, Chandigarh, India}
\affiliation{Delhi University, Delhi, India}
\affiliation{Tata Institute of Fundamental Research, Mumbai, India}
\affiliation{University College Dublin, Dublin, Ireland}
\affiliation{Korea Detector Laboratory, Korea University, Seoul, Korea}
\affiliation{CINVESTAV, Mexico City, Mexico}
\affiliation{Nikhef, Science Park, Amsterdam, the Netherlands}
\affiliation{Radboud University Nijmegen, Nijmegen, the Netherlands}
\affiliation{Joint Institute for Nuclear Research, Dubna, Russia}
\affiliation{Institute for Theoretical and Experimental Physics, Moscow, Russia}
\affiliation{Moscow State University, Moscow, Russia}
\affiliation{Institute for High Energy Physics, Protvino, Russia}
\affiliation{Petersburg Nuclear Physics Institute, St. Petersburg, Russia}
\affiliation{Instituci\'{o} Catalana de Recerca i Estudis Avan\c{c}ats (ICREA) and Institut de F\'{i}sica d'Altes Energies (IFAE), Barcelona, Spain}
\affiliation{Uppsala University, Uppsala, Sweden}
\affiliation{Taras Shevchenko National University of Kyiv, Kiev, Ukraine}
\affiliation{Lancaster University, Lancaster LA1 4YB, United Kingdom}
\affiliation{Imperial College London, London SW7 2AZ, United Kingdom}
\affiliation{The University of Manchester, Manchester M13 9PL, United Kingdom}
\affiliation{University of Arizona, Tucson, Arizona 85721, USA}
\affiliation{University of California Riverside, Riverside, California 92521, USA}
\affiliation{Florida State University, Tallahassee, Florida 32306, USA}
\affiliation{Fermi National Accelerator Laboratory, Batavia, Illinois 60510, USA}
\affiliation{University of Illinois at Chicago, Chicago, Illinois 60607, USA}
\affiliation{Northern Illinois University, DeKalb, Illinois 60115, USA}
\affiliation{Northwestern University, Evanston, Illinois 60208, USA}
\affiliation{Indiana University, Bloomington, Indiana 47405, USA}
\affiliation{Purdue University Calumet, Hammond, Indiana 46323, USA}
\affiliation{University of Notre Dame, Notre Dame, Indiana 46556, USA}
\affiliation{Iowa State University, Ames, Iowa 50011, USA}
\affiliation{University of Kansas, Lawrence, Kansas 66045, USA}
\affiliation{Louisiana Tech University, Ruston, Louisiana 71272, USA}
\affiliation{Northeastern University, Boston, Massachusetts 02115, USA}
\affiliation{University of Michigan, Ann Arbor, Michigan 48109, USA}
\affiliation{Michigan State University, East Lansing, Michigan 48824, USA}
\affiliation{University of Mississippi, University, Mississippi 38677, USA}
\affiliation{University of Nebraska, Lincoln, Nebraska 68588, USA}
\affiliation{Rutgers University, Piscataway, New Jersey 08855, USA}
\affiliation{Princeton University, Princeton, New Jersey 08544, USA}
\affiliation{State University of New York, Buffalo, New York 14260, USA}
\affiliation{University of Rochester, Rochester, New York 14627, USA}
\affiliation{State University of New York, Stony Brook, New York 11794, USA}
\affiliation{Brookhaven National Laboratory, Upton, New York 11973, USA}
\affiliation{Langston University, Langston, Oklahoma 73050, USA}
\affiliation{University of Oklahoma, Norman, Oklahoma 73019, USA}
\affiliation{Oklahoma State University, Stillwater, Oklahoma 74078, USA}
\affiliation{Brown University, Providence, Rhode Island 02912, USA}
\affiliation{University of Texas, Arlington, Texas 76019, USA}
\affiliation{Southern Methodist University, Dallas, Texas 75275, USA}
\affiliation{Rice University, Houston, Texas 77005, USA}
\affiliation{University of Virginia, Charlottesville, Virginia 22904, USA}
\affiliation{University of Washington, Seattle, Washington 98195, USA}
\author{V.M.~Abazov} \affiliation{Joint Institute for Nuclear Research, Dubna, Russia}
\author{B.~Abbott} \affiliation{University of Oklahoma, Norman, Oklahoma 73019, USA}
\author{B.S.~Acharya} \affiliation{Tata Institute of Fundamental Research, Mumbai, India}
\author{M.~Adams} \affiliation{University of Illinois at Chicago, Chicago, Illinois 60607, USA}
\author{T.~Adams} \affiliation{Florida State University, Tallahassee, Florida 32306, USA}
\author{J.P.~Agnew} \affiliation{The University of Manchester, Manchester M13 9PL, United Kingdom}
\author{G.D.~Alexeev} \affiliation{Joint Institute for Nuclear Research, Dubna, Russia}
\author{G.~Alkhazov} \affiliation{Petersburg Nuclear Physics Institute, St. Petersburg, Russia}
\author{A.~Alton$^{a}$} \affiliation{University of Michigan, Ann Arbor, Michigan 48109, USA}
\author{A.~Askew} \affiliation{Florida State University, Tallahassee, Florida 32306, USA}
% ...
%
%
\collaboration{The D0 Collaboration\footnote{with visitors from
$^{a}$Augustana College, Sioux Falls, SD, USA,
$^{b}$The University of Liverpool, Liverpool, UK,
$^{c}$DESY, Hamburg, Germany,
$^{d}$Universidad Michoacana de San Nicolas de Hidalgo, Morelia, Mexico
$^{e}$SLAC, Menlo Park, CA, USA,
$^{f}$University College London, London, UK,
$^{g}$Centro de Investigacion en Computacion - IPN, Mexico City, Mexico,
$^{h}$Universidade Estadual Paulista, S\~ao Paulo, Brazil,
$^{i}$Karlsruher Institut f\"ur Technologie (KIT) - Steinbuch Centre for Computing (SCC),
D-76128 Karlsruhe, Germany,
$^{j}$Office of Science, U.S. Department of Energy, Washington, D.C. 20585, USA,
$^{k}$American Association for the Advancement of Science, Washington, D.C. 20005, USA,
$^{l}$Kiev Institute for Nuclear Research, Kiev, Ukraine
and
$^{m}$University of Maryland, College Park, Maryland 20742, USA.
}} \noaffiliation
\vskip 0.25cm

\date{May 15, 2014}

\begin{abstract}
  We present the first measurements of the differential cross section ${\rm d}\sigma/{\rm d}\Ptg$
  for the production of an isolated photon in association with at least two $b$-quark jets.
  The measurements consider photons with rapidities $|y^\gamma|\lt 1.0$ and transverse momenta   
  $30<\ptg <200$~\GeV. The $b$-quark jets are required to have $p_T^\text{jet}>15$ GeV and $| y^\text{jet}|\lt 1.5$.
  The ratio of differential production cross sections for $\gamma+2~b$-jets to $\gamma+b$-jet as a function of \ptg 
  is also presented.
  The results are based on the proton-antiproton collision data at $\sqrt{s}=$1.96~\TeV collected with the
  D0 detector at the Fermilab Tevatron Collider. The measured cross sections and their ratios are compared to the 
  next-to-leading order perturbative QCD calculations as well as  predictions
  based on the $k_{\rm T}$-factorization approach and those from the {\sc sherpa} and {\sc pythia} Monte Carlo event generators.

\end{abstract}

\maketitle

In hadronic collisions, high-energy photons ($\gamma$)
emerge unaltered from the hard parton-parton interaction and
therefore provide a clean probe of the underlying hard-scattering dynamics~\cite{Owens}. Photons produced in these interactions (called direct or prompt) in association with one or more bottom ($b$)-quark jets provide an 
important test of perturbative Quantum Chromodynamics (QCD) predictions at large hard-scattering scales $Q$ and
over a wide range of parton momentum fractions. In addition, the study of these
processes also provides information about the parton density functions (PDF) of $b$ quarks and gluons ($g$), which still have substantial uncertainties.
In $p\bar{p}$ collisions, \gb-jet events are produced primarily through the 
Compton process $gb\to \gamma b$, which dominates for low and moderate photon transverse momenta ($\Ptg$), and 
through quark-antiquark annihilation followed by $g \to b\bar{b}$ gluon splitting $q\bar{q}\to \gamma g \to \gamma b\bar{b}$, which dominates at high $\Ptg$~\cite{Tzvet,gamma_b_d0_2}.
The final state with $b$-quark pair production, $p\bar{p} \rightarrow \gamma+b\bar{b}$, is
mainly produced via $q\bar{q}\to \gamma b\bar{b}$ and $gg\to \gamma b\bar{b}$ scatterings~\cite{Reina}.
The $\gamma+2~b$-jet process is a crucial component of background 
in measurements of, for example, $t\bar{t} \gamma$ coupling~\cite{ttgam} and in some searches for new phenomena. 
A series of measurements involving
$\gamma$ and $b(c)$-quark final states have previously been performed by the D0 and 
CDF Collaborations~\cite{gamma_b_d0_1,gamma_b_d0_2, gamma_b_d0_3, gamma_b_cdf_1, gamma_b_cdf_2}.

In this measurement, we follow an inclusive approach by allowing 
the final state with any additional jet(s) on top of the studied $b$-quark jets. Inclusive $\gamma+2~b$-jet production may also originate from partonic subprocesses
involving parton fragmentation into a photon. However, using photon isolation requirements
significantly reduces the contributions from such processes.
Next-to-leading order (NLO) calculations of the $\gamma+2~b$-jet production cross section, which
includes all $b$-quark mass effects, have recently become available~\cite{Reina}.
These calculations are based on the four-flavor number scheme, which assumes 
four massless quark flavors and treats the $b$ quark as a massive quark not appearing in the initial state.

This letter presents the first measurement of the cross section for
associated production of an isolated photon with a bottom quark pair in $p\bar{p}$ collisions. The results are based on data corresponding
to an integrated luminosity of $8.7 \pm 0.5$~fb$^{-1}$~\cite{d0lumi} collected with the D0 detector from June 2006 to September 
2011 at the Fermilab Tevatron Collider at $\sqrt{s}=$1.96~\TeV.
The large data sample and use of advanced photon and $b$-jet identification tools~\cite{diphoton,hgg_prl,b-NIM}
enable us to measure the $\gamma+2~b$-jet production cross section differentially 
as a function of $\Ptg$ for photons with rapidities $|y^\gamma|\lt 1.0$ and transverse momenta $30<\Ptg< 200$~GeV, while the 
$b$ jets are required to have $p_{T}^{\rm jet}>15$~GeV and $| y^\text{jet}|\lt 1.5$. 
This allows for probing the dynamics of the production process over
a wide kinematic range not studied before in other measurements of 
a vector boson + $b$-jet final state. 
The ratio of differential cross
sections for $\gamma+2~b$-jet production relative to $\gamma+b$-jet production is also
presented in the same kinematic region and differentially in $\Ptg$. The measurement of the ratio of cross sections leads to cancellation
of various experimental and theoretical uncertainties, allowing a more precise comparison with the theoretical predictions.

The D0 detector is a general purpose detector 
described in detail elsewhere~\cite{d0det}.
The subdetectors most relevant to this analysis are the central tracking
system, composed of a silicon microstrip tracker (SMT) and a central fiber
tracker embedded in a 1.9~T solenoidal magnetic field, the central
preshower detector (CPS), and the calorimeter.
The CPS is located immediately before the inner layer of the central calorimeter
and is formed of approximately one radiation length of lead absorber followed by three
layers of scintillating strips. The calorimeter consists of a central section (CC) with
coverage in pseudorapidity of $|\eta_{\rm det}|<1.1$~\cite{d0_coordinate}, 
and two end calorimeters (EC) extending coverage to $|\eta_{\rm det}| \approx 4.2$, each housed
in a separate cryostat, with scintillators between the CC and EC cryostats providing sampling of
developing showers for $1.1 \lt |\eta_{\rm det}| \lt 1.4$.
The electromagnetic (EM) section of the
calorimeter is segmented longitudinally into four layers (EM$i$, $i=1-4$), 
with transverse segmentation into cells of size 
$\Delta\eta_{\rm det}\times\Delta\phi_{\rm det} = 0.1\times 0.1$~\cite{d0_coordinate}, 
except EM3 (near the EM shower maximum), where it is $0.05\times 0.05$.
The calorimeter allows for a precise measurement of the energy of electrons and photons,
providing an energy resolution of approximately $4\%$~($3\%$) at an energy of $30~(100)$~GeV.
The energy response of the calorimeter to photons is calibrated using
electrons from $Z$ boson decays. Because electrons and photons interact
differently in the detector material before the calorimeter, additional energy corrections 
as a function of $\Ptg$ are derived using a detailed {\sc geant}-based~\cite{Geant} simulation 
of the D0 detector response. These corrections are $\approx 2$\% for photon candidates of $\Ptg = 30$ GeV, and smaller for higher $\Ptg$.

The data used in this analysis satisfy D0 experiment data quality requirements and
are collected using a combination of triggers 
requiring a cluster of energy in the EM calorimeter with
loose shower shape requirements. The trigger efficiency is $\approx\!96\%$ for photon candidates with 
$p_T^\gamma=30$~GeV and $100\%$ for $p_T^\gamma \gtrsim 40$~GeV.
Offline event selection requires a reconstructed $p\bar{p}$ interaction vertex~\cite{pv} 
within 60~cm of the center of the detector along the beam axis. 
The efficiency of the vertex requirement
is $\approx\!(96-98)\%$, depending on $\Ptg$. The missing transverse momentum in the event is 
required to be less than $0.7 p_{T}^{\gamma}$ to suppress background from 
$W\to e\nu$ decays.  Such a requirement is highly efficient ($\geq 98\%$) for signal events.

The photon selection criteria in the current measurement are identical to those used in Refs.~\cite{gamma_b_d0_1, gamma_b_d0_2}.
The photon selection efficiency and acceptance
are calculated using samples of $\gamma+b$-jet events,
generated with the {\sc sherpa}~\cite{Sherpa} and {\sc pythia} \cite{PYT} Monte Carlo (MC) event generators. 
The samples are  processed through a {\sc geant}-based~\cite{Geant} 
simulation of the D0 detector. Simulated events are overlaid with data events from random $p\bar{p}$ crossings to properly model the effects of multiple $p\bar{p}$ interactions and noise in data. We ensure that the instantaneous luminosity distribution in the overlay events is similar to the data. The efficiency for photons to pass the identification criteria is ($71-82$)\% with relative systematic uncertainty of $3$\%.

For the $\gamma+n~b$ measurement ($n=1,2$), $n$ jets with the highest $p_{T}$ that satisfy
$p_{T}^{\rm jet}>15$~GeV and $| y^\text{jet}|\lt 1.5$ are selected. Jets are reconstructed
using the D0 Run~II algorithm~\cite{Run2Cone} with a cone radius of $\mathcal{R}=0.5$.  
A set of criteria is imposed to ensure that we have sufficient information to 
identify the jet as a heavy-flavor candidate:
the jet is required to have at least two associated tracks with $p_T>0.5$~\GeV
and at least one hit in the SMT, one of these tracks must also have $p_T>1.0$~\GeV.
These criteria 
have an efficiency of about 90\% for a $b$ jet.  Light jets (initiated by $u$, $d$ and $s$ quarks  or gluons) are suppressed
using a dedicated heavy-flavor (HF) tagging algorithm ~\cite{b-NIM}. 

The HF tagging algorithm is based on a multivariate analysis (MVA) technique that
combines information from the secondary vertex (SV) tagging algorithms and
tracks impact parameter variables using an artificial neural network (NN) to define a single output discriminant, MVA$_{\rm bl}$~\cite{b-NIM}.
This algorithm utilizes the longer
lifetimes of HF hadrons relative to their lighter counterparts.
The MVA$_{\rm bl}$ has a continuous output value that tends towards one
for $b$ jet and zero for light jets.
Events with at least two jets
passing the MVA$_{\rm bl} > 0.3$ selection are considered in the $\gamma + 2~b$-jet analysis.
Depending on $\ptg$, this selection has an efficiency of ($13-21$)\% for two $b$ jets
with relative systematic uncertainties of ($4-6$)\%, primarily due to uncertainties on the data-to-MC correction factors~\cite{b-NIM}.
Only ($0.2-0.4$)\% of light-jets are misidentified as $b$ jets.

\begin{figure}
\includegraphics[width=1.03\linewidth]{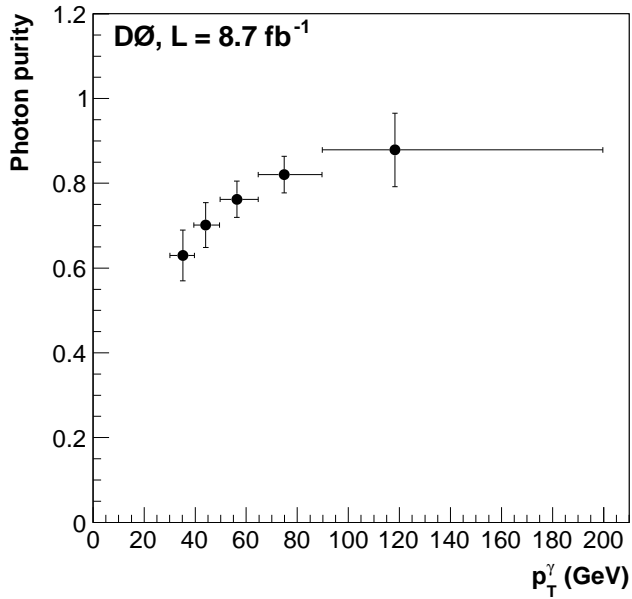} 
~\\[-4mm]
\caption{Photon purity 
as a function of $\Ptg$ in the selected data sample. The error bars include statistical and systematic uncertainties added in quadrature.
The binning is defined as in Table \ref{tab:xsect_gbb}.}
\label{fig:phot_pur}
\end{figure}

After application of all selection requirements, 3,816 $\gamma+2~b$-jet candidate (186,406 $\gamma+b$-jet candidate) events remain in the data sample.
In these events, there are two main background sources: jets misidentified as photons and light-flavor jets mimicking HF jets. To estimate the
photon purity, the $\gamma$-NN distribution in data is fitted to a
linear combination of templates for photons and jets obtained from
simulated $\gamma~+$ jet and dijet samples.  An
independent fit is performed in each $\Ptg$ bin, yielding photon
fractions between 62\% and 90\%,
as shown in Fig.~\ref{fig:phot_pur}. The main systematic uncertainty in the photon fractions is due to
 the fragmentation model implemented in {\sc pythia}~\cite{PhotonInc}.
This uncertainty is estimated by varying the
production rate of $\pi^0$ and $\eta$ mesons by $\pm$50\% with respect to their
central values~\cite{Frag}, and found to be about $6\%$ at $\Ptg \approx 30$ GeV, 
and $\leq1\%$ at $\Ptg\gtrsim 70$ GeV.

\begin{figure}
\includegraphics[width=1.03\linewidth]{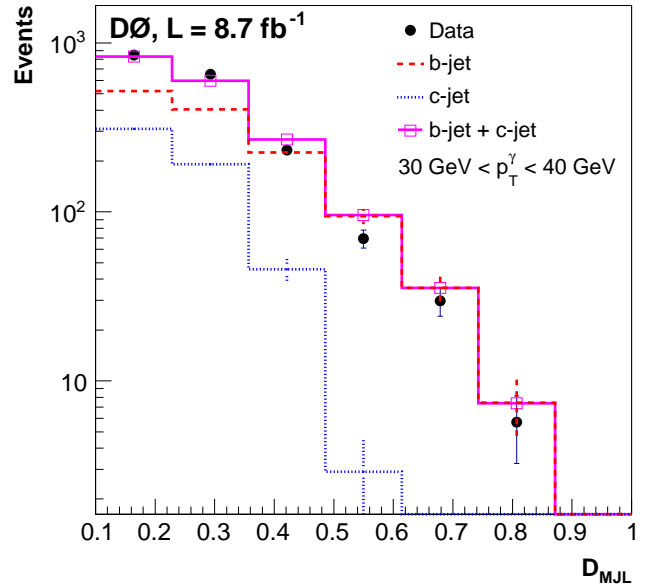} 
\caption{(Color online) Distribution of discriminant $D_{\rm MJL}$
  after all selection criteria for a representative bin of $30<\Ptg< 40$~GeV.  
  The expected contribution from the light jets component has been subtracted from the data.
  The distributions for the $b$-jet and $c$-jet templates (with statistical uncertainties) 
  are shown normalized to their respective fitted fractions.
  }
\label{fig:cbjet_test}
\end{figure}

\begin{figure}
\includegraphics[width=1.03\linewidth]{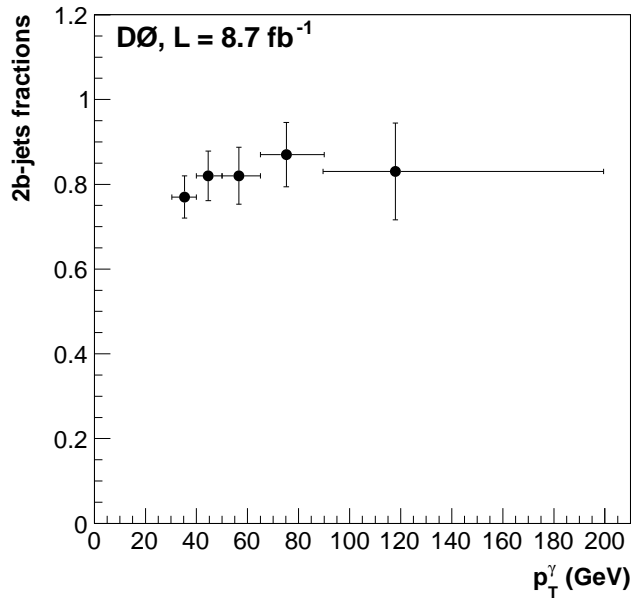} 
~\\[-4mm]
\caption{The $2~b$-jet fraction in data 
as a function of $\Ptg$ derived from the template fit to the heavy quark jet data sample
after applying all selections. The error bars show both statistical and systematical uncertainties summed in quadrature.
Binning is the same as given in Table \ref{tab:xsect_gbb}.}
\label{fig:b_pur_ptg}
\end{figure}
The fraction of different flavor jets in the
selected data sample is extracted using a discriminant, $D_{\rm MJL}$,
with distributions dependent on  the jet flavors.
It combines two discriminating variables associated with the jet, mass of any secondary vertex associated with the jet $M_{\rm SV}$  and the
probability for the jet tracks located within the jet cone to come from the primary $p\bar{p}$ interaction vertex.
The latter probability is found using the jet lifetime impact parameter (JLIP) algorithm,
and is denoted as $P_{\rm JLIP}$~\cite{pv}. The final $D_{\rm MJL}$ discriminant~\cite{DMJL_wb} is defined as
$D_{\rm MJL}=0.5 \times (M_{\rm SV}/{5~\rm GeV}-\ln(P_{\rm JLIP})/20)$, where
$M_{\rm SV}$ and $\ln(P_{\rm JLIP})$ are normalized by their maximum values obtained from the corresponding distributions in data.
The data sample with two HF-tagged jets is fitted to templates 
 consisting mainly of $2~b$-jet and $2~c$-jet events, as determined from MC simulation.
The remaining jet flavor contributions in the sample (e.g., light+light-jets, light+$b(c)$-jets, etc)
are small and are subtracted from the data. 
The fractions of these rarer jet contributions are estimated from {\sc sherpa} simulation
(which has been found to provide a good description of the data), and vary in the range ($5-10$)\%.
The difference in the values of these fractions obtained from {\sc sherpa} and {\sc pythia}, ($2-4$)\%,
is assigned as a systematic uncertainty on the background estimate.
The fraction of $2~b$-jet events are determined by performing a two-dimensional
(corresponding to the $2~b$-jet candidates) maximum likelihood fit of $D_{\rm MJL}$
distributions of $2$ jet events in data using the corresponding templates for $2~b$-jets and $2~c$-jets.
These jet flavor templates are obtained from MC simulations.
As an example, the result of one of these maximum likelihood fits to $D_{\rm MJL}$ templates
is presented in Fig.~\ref{fig:cbjet_test} (with
$\chi^2/ndf=6.80/5$ for data/MC agreement).
This shows the one-dimensional projection onto the highest $p_T$ jet $D_{\rm MJL}$ axis of the 2D fit, normalized to the number of events in data, for photons with $30<\Ptg<40~\GeV$. An independent fit is performed in each $\ptg$ bin, resulting in extracted fractions of $2~b$-jet events
between $76$\% and $87$\%, as shown in Fig.~\ref{fig:b_pur_ptg}.
The relative uncertainties of the estimated $2~b$-jet fractions range from
$5$\% to $14$\%, increasing at higher $\Ptg$ and are dominated by the limited data statistics.

\begin{figure}
\includegraphics[width=1.03\linewidth]{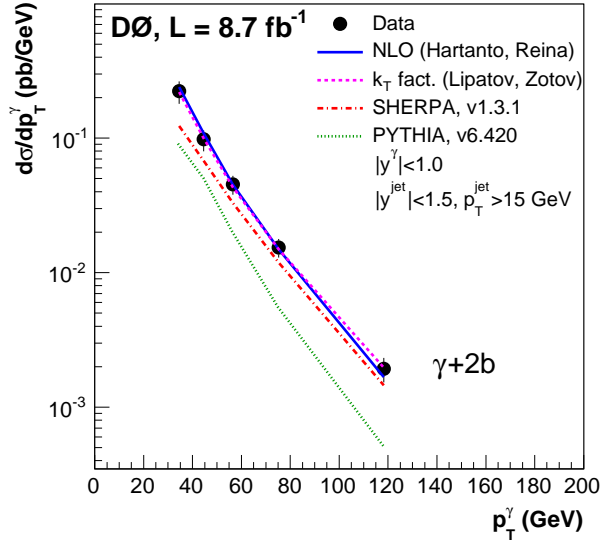}
~\\[-4mm]
\caption{(Color online) The $\gamma+2~b$-jet differential production cross sections
  as a function of $\Ptg$.
 The uncertainties on the data points include statistical and systematic contributions. 
 The measurements are compared to the NLO QCD calculations~\cite{Reina} using the {\sc cteq}6.6M PDFs~\cite{CTEQ} (solid line). The predictions from {\sc sherpa}~\cite{Sherpa}, {\sc pythia}~\cite{PYT} and the $k_{\rm T}$-factorization approach~\cite{Zotov,Zotov2} are also shown.} 
\label{fig:xsectLOGplot}
\end{figure}

\begin{figure}
\includegraphics[width=1.03\linewidth]{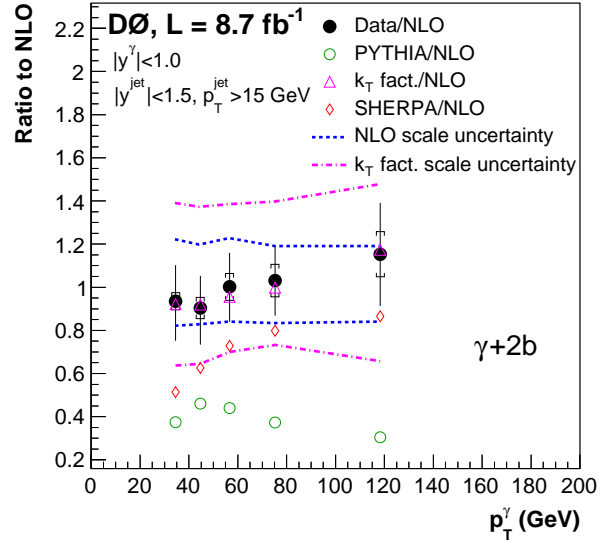}
~\\[-4mm]
\caption{(Color online) The ratio of the measured $\gamma+2~b$-jet production cross sections to the reference NLO predictions.
The uncertainties on the data include both statistical (inner error bar) and total uncertainties (full error bar).  
Similar ratios to NLO calculations for predictions with {\sc sherpa}~\cite{Sherpa}, {\sc pythia}~\cite{PYT} and $k_{\rm T}$-factorization \cite{Zotov,Zotov2} are also presented along with the scale uncertainties on NLO and $k_{\rm T}$-factorization predictions.
}
\label{fig:xsectratio}
\end{figure}
The estimated numbers of signal events in each $\Ptg$ bin are
corrected for the geometric and kinematic acceptance
of the photon and jets. The combined acceptance for photon and jets  are calculated using {\sc sherpa} MC events. The acceptance is calculated for
the photons satisfying  $\Ptg > 30$ GeV, $|y^\gamma|\lt 1.0$ at particle level. The particle level includes all stable particles as defined in Ref.~\cite{particle}. The jets are required to have $p_T^\text{jet}>15$ GeV and $| y^\text{jet}|\lt 1.5$.
As in Refs.~\cite{gamma_b_d0_1, gamma_b_d0_2},
in the acceptance calculations, the photon is required to be isolated 
by $E_T^{\rm iso} = E_T^{\rm tot}(0.4) - E_T^\gamma < 2.5$ GeV,
where $E_T^{\rm tot}(0.4)$ is the total transverse energy of particles within a cone of radius ${\cal R} = \sqrt{(\Delta\eta)^2+(\Delta\phi)^2} = 0.4$
centered on the photon direction and $E_T^\gamma$ is the photon transverse energy.
The sum of transverse energy in the cone includes all stable particles.~\cite{particle}.
The acceptance is driven by selection requirements in $|\eta_{\rm det}|$
(applied to avoid edge effects in the calorimeter regions used for the measurement) and
$|\phi_{\rm det}|$ (to avoid periodic calorimeter module boundaries), photon $|\eta^\gamma|$
and $\Ptg$, and bin-to-bin migration effects due to the finite energy resolution of the
EM calorimeter. The combined photon and jets acceptance with respect to the $p_{T}$ and rapidity selections
varies between $66\%$ and $77\%$ in different $p_T^\gamma$ bins.
Uncertainties on the acceptance due to the jet energy scale~\cite{JES}, jet energy
resolution, and the difference between results obtained with {\sc sherpa} and {\sc pythia} are in the range of ($8-12$)\%.

The data, corrected for photon and jet acceptance, reconstruction efficiencies
and the admixture of background events, are presented at the particle level  
by unfolding for effects of detector resolution, photon and $b$-jet detection inefficiencies.
The differential cross sections of $\gamma+2~b$-jet production are extracted in five bins of \ptg{}. 
They are given in Table~\ref{tab:xsect_gbb}. 
The data points are plotted at the values of $\Ptg$ for which
the value of a smooth function describing the dependence of the cross section on \ptg
equals the averaged cross section in the bin~\cite{TW}.
 
The cross sections fall by
more than two orders of magnitude in the range $30<\Ptg< 200~\GeV$.
The statistical uncertainty on the results ranges from 4.3\% in the
first $\Ptg$ bin to $9\%$ in the last $\Ptg$ bin, while the total
systematic uncertainty ranges up to 20\%.  
Main sources of systematic
uncertainty are the photon purity (up to $8\%$), photon and two $b$-jet acceptance (up to $14\%$),
$b$-jet fraction (up to $13\%$), and integrated luminosity ($6$\%)~\cite{d0lumi}.  
At higher \ptg, the uncertainty is
dominated by the fractions of $b$-jet events and their selection efficiencies.

NLO perturbative QCD predictions, with
the renormalization scale $\mu_{R}$, factorization scale $\mu_{F}$,
and fragmentation scale $\mu_f$ all set to $\Ptg$, are also given in
Table \ref{tab:xsect_gbb}.
The uncertainty from the scale choice is ($15-20$)\% and is estimated 
through a simultaneous variation of all three scales by a factor of two, i.e., 
 for $\mu_{R,F,f}=0.5 p_T^\gamma$ and $2 p_T^\gamma$. 
The predictions utilize {\sc cteq}6.6M PDFs~\cite{CTEQ} and are corrected for non-perturbative effects
of parton-to-hadron fragmentation and multiple parton interactions. 
The latter are evaluated using {\sc sherpa} and {\sc pythia} MC samples with their standard
settings~\cite {Sherpa,PYT}.
The overall correction varies from about $0.90$ at $30<\Ptg<40~\GeV$ to about $0.95$ at high $\Ptg$,
and an uncertainty of $\lesssim 2\%$ is assigned to account for 
differences between the two MC generators.

The predictions based on the $k_T$-factorization approach~\cite{Zotov,Zotov2} 
and unintegrated parton distributions~\cite{UPD}
are also given in Table \ref{tab:xsect_gbb}.
The $k_T$-factorization formalism contains additional contributions 
to the cross sections due to resummation of
gluon radiation diagrams with $k_T^2$ above a scale $\mu^2$ of ${\cal{O}}(1~ \rm{GeV})$, 
where  $k_T$ denotes the transverse momentum of the radiated gluon. 
Apart from this resummation, the non-vanishing transverse momentum 
distribution of the colliding partons are taken into account. 
These effects lead to a broadening of the photon transverse momentum distribution 
in this approach~\cite{Zotov}.
The scale uncertainties on these predictions 
vary from about $31\%$ at $30<\Ptg<40~\GeV$ to about 
50\% in the highest $\Ptg$ bin.

Table \ref{tab:xsect_gbb} also contains predictions 
from the {\sc pythia} \cite{PYT} MC event generator with the {\sc cteq}6.1L PDF set.
It includes only $2\to 2$ matrix elements (ME) with $gb\to \gamma b$ and $q\bar{q}\to \gamma g$ scatterings 
(defined at LO) and with $g\to b\bar{b}$ splitting in the parton shower (PS).
We also provide predictions of the
{\sc sherpa} MC event generator \cite{Sherpa} with the {\sc cteq}6.6M  PDF set~\cite{CTEQ}.
For \gb production, {\sc sherpa}
includes all the MEs with one photon and up to three
jets, with at least one $b$-jet in our kinematic region. 
In particular, 
it accounts for an additional hard jet that accompanies the photon 
associated with $2~b$ jets. Compared to an  NLO calculation,
there is an additional benefit of imposing resummation (further emissions)
through the consistent combination with the PS.
Matching between the ME partons and the PS jets follows the
prescription given in Ref.~\cite{SHERPA_gam}.
Systematic uncertainties are estimated by varying the ME-PS matching scale by $\pm 5$ GeV
around the chosen central value \cite{Sherpa_scale}.
As a result, the {\sc sherpa} cross sections vary up to $\pm7\%$, the uncertainty being largest in the first $\Ptg$ bin.

All the theoretical predictions are obtained including the isolation requirement 
on the photon $E_T^{\rm iso}<2.5$ GeV.
The predictions are compared to data in Fig.~\ref{fig:xsectLOGplot} as a function of $\Ptg$.
The ratios of data to the NLO QCD calculations and of different QCD predictions or MC simulation to
the same NLO QCD calculations are shown in Fig.~\ref{fig:xsectratio} as a function of $\Ptg$.

\begin{figure}
\includegraphics[width=1.03\linewidth]{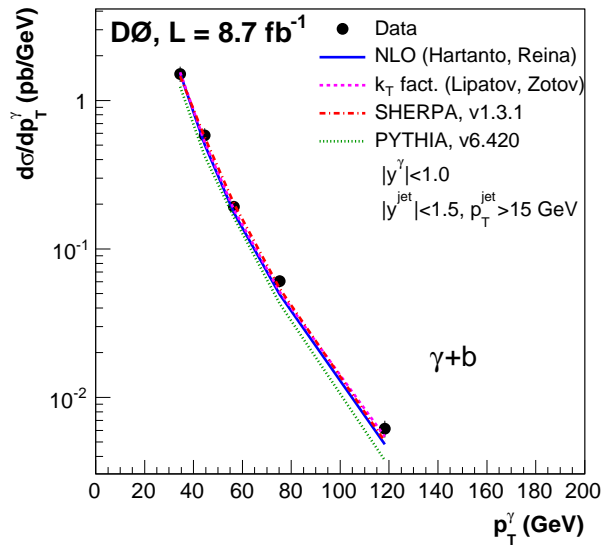}
~\\[-4mm]
\caption{(Color online) The $\gamma+b$-jet differential production cross sections
  as a function of $\Ptg$. The uncertainties on the data points include statistical and systematic contributions added in quadrature.
The measurements are compared to the NLO QCD calculations~\cite{Reina} using the {\sc cteq}6.6M
  PDFs~\cite{CTEQ} (solid line).
  The predictions from {\sc sherpa}~\cite{Sherpa}, {\sc pythia}~\cite{PYT} and  $k_{\rm T}$-factorization \cite{Zotov,Zotov2} 
  are also shown.}
\label{fig:xsectLOGplot_gb}
\end{figure}

\begin{figure}
\includegraphics[width=1.03\linewidth]{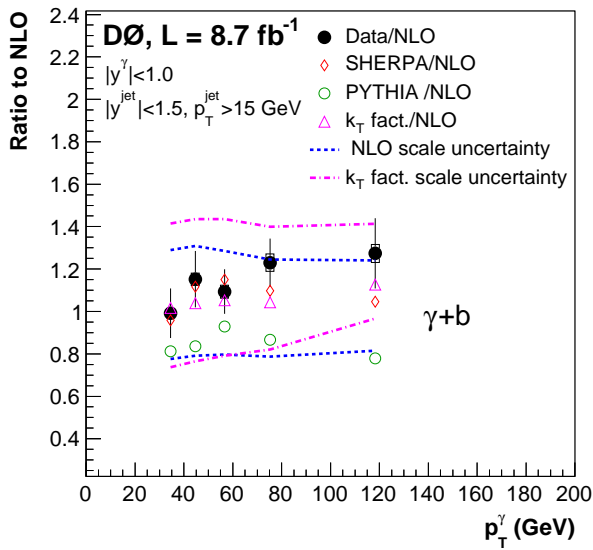}
~\\[-4mm]
\caption{(Color online) 
The ratio of $\gamma+b$-jet production cross sections to NLO predictions for data and theoretical predictions.
The uncertainties on the data include both statistical (inner error bar) and total uncertainties (full error bar).  
The ratios to the NLO calculations with predictions from {\sc sherpa}~\cite{Sherpa}, {\sc pythia}~\cite{PYT} and $k_{\rm T}$-factorization \cite{Zotov,Zotov2} are also presented along with the scale uncertainties on NLO and $k_{\rm T}$-factorization predictions.
}
\label{fig:xsectratio_gb}
\end{figure}

\begin{table*}
\centering
\caption{The differential $\gamma+2~b$-jet production cross sections ${\rm d}\sigma/{\rm d}\Ptg$ in bins of $\Ptg$ for $|\eta^\gamma|<1.0$, $p_T^\text{jet}>15$ GeV and $| y^\text{jet}|\lt 1.5$ together with statistical uncertainties ($\delta_{\rm stat}$), total systematic uncertainties ($\delta_{\rm syst}$) and total uncertainties ($\delta_{\rm tot}$) which are obtained by adding $\delta_{\rm stat}$ and $\delta_{\rm syst}$ in quadrature. The last four columns show theoretical predictions obtained with NLO QCD, $k_{\rm T}$ factorization, and with the {\sc pythia} and the {\sc sherpa} event generators. }
\label{tab:xsect_gbb}
\begin{tabular}{cccccccccc} 
\hline \hline
~$\Ptg$ bin~ & ~$\la\Ptg\ra$~ & \multicolumn{8}{c}{${\rm d}\sigma/{\rm d}\Ptg$ (pb/GeV) } \\\cline{3-10}
  (GeV) & (GeV) & Data & $\delta_{\rm stat}$($\%$) & $\delta_{\rm syst}$($\%$) & $\delta_{\rm tot}$($\%$) & ~~~NLO~~~ & ~~~$k_{\rm T}$ fact.~~~ & ~~~{\sc pythia}~~~ & ~~~{\sc sherpa}~~~\\\hline

30 --  40 &   34.5 &  2.24$\times 10^{-1}$& 4.3 & +19/$-17$~~ & +19/$-18$~~& 2.39$\times 10^{-1}$~~& 2.20$\times 10^{-1}$~~& 8.96$\times 10^{-2}$~~& 1.23$\times 10^{-1}$\\
40 --  50 &   44.6 &  9.80$\times 10^{-2}$ & 5.4 & +18/$-15$~~ & +19/$-16$~~& 1.08$\times 10^{-1}$~~& 9.96$\times 10^{-2}$~~& 4.99$\times 10^{-2}$~~& 6.79$\times 10^{-2}$\\ 
50 --  65 &   56.6 &  4.52$\times 10^{-2}$ & 6.2 & +15/$-14$~~ & +16/$-16$~~& 4.51$\times 10^{-2}$~~& 4.31$\times 10^{-2}$~~& 1.99$\times 10^{-2}$~~& 3.29$\times 10^{-2}$\\
65 --  90 &   75.2 &  1.54$\times 10^{-2}$ & 7.2 & +14/$-14$~~ & +16/$-16$~~& 1.49$\times 10^{-2}$~~& 1.48$\times 10^{-2}$~~& 5.57$\times 10^{-3}$~~& 1.19$\times 10^{-2}$\\
90 -- 200 &  118.3 &  1.93$\times 10^{-3}$ & 9.1 & +19/$-18$~~ & +21/$-21$~~& 1.67$\times 10^{-3}$~~& 1.96$\times 10^{-3}$~~& 5.12$\times 10^{-4}$~~& 1.45$\times 10^{-3}$\\
\hline \hline
\end{tabular}
\end{table*}

\begin{table*}
\centering
\caption{The differential $\gamma+b$-jet production cross sections ${\rm d}\sigma/{\rm d}\Ptg$ in bins of $\Ptg$ for $|\eta^\gamma|<1.0$, $p_T^\text{jet}>15$ GeV and $| y^\text{jet}|\lt 1.5$ together with statistical uncertainties ($\delta_{\rm stat}$), total systematic uncertainties ($\delta_{\rm syst}$), and total uncertainties ($\delta_{\rm tot}$) that are obtained by adding $\delta_{\rm stat}$ and $\delta_{\rm syst}$ in quadrature. The last four columns show theoretical predictions obtained with NLO QCD, $k_{\rm T}$-factorization, and with the {\sc pythia} and the {\sc sherpa} event generators.}
\label{tab:xsect_gb}
\begin{tabular}{cccccccccc}
\hline \hline
~$\Ptg$ bin~ & ~$\la\Ptg\ra$~ & \multicolumn{8}{c}{${\rm d}\sigma/{\rm d}\Ptg$ (pb/GeV) } \\\cline{3-10}
  (GeV) & (GeV) & Data & $\delta_{\rm stat}$($\%$) & $\delta_{\rm syst}$($\%$) & $\delta_{\rm tot}$($\%$) & ~~~NLO~~~ & ~~~$k_{\rm T}$ fact.~~~ & ~~~{\sc pythia}~~~ & ~~~{\sc sherpa}~~~\\\hline

 30 --  40 &   34.5 & 1.51 &2.3 & 12 & 12&1.52& 1.69&1.23&1.46 \\
  40 --  50 &   44.6 &  5.83$\times 10^{-1}$ & 2.4 & 11 & 12&5.06$\times 10^{-1}$&5.70$\times 10^{-1}$&4.23$\times 10^{-1}$&5.65$\times 10^{-1}$\\
  50 --  65 &   56.6 & 1.92$\times 10^{-1}$ & 2.8 & 9 & 10&1.75$\times 10^{-1}$&1.98$\times 10^{-1}$& 1.63$\times 10^{-1}$&2.02$\times 10^{-1}$\\
  65 --  90 &   75.2 & 6.06$\times 10^{-2}$ & 3.3 & 9 & 9&4.93$\times 10^{-2}$&5.43$\times 10^{-2}$&4.27$\times 10^{-2}$&5.41$\times 10^{-2}$\\
  90 -- 200 &   118.3 &  6.15$\times 10^{-3}$ & 3.3& 13 & 13&4.83$\times 10^{-3}$& 5.68$\times 10^{-3}$&3.76$\times 10^{-3}$&5.05$\times 10^{-3}$\\ 
\hline \hline
\end{tabular}
\end{table*}

\begin{table*}

\centering
\caption{The $\sigma$($\gamma+2~b$-jet)/$\sigma$($\gamma+b$-jet) cross section ratio in bins of $\Ptg$ for $|\eta^\gamma|<1.0$, $p_T^\text{jet}>15$ GeV and $| y^\text{jet}|\lt 1.5$ together with statistical uncertainties ($\delta_{\rm stat}$), total systematic uncertainties ($\delta_{\rm syst}$) and total uncertainties ($\delta_{\rm tot}$) which are obtained by adding $\delta_{\rm stat}$ and $\delta_{\rm syst}$ in quadrature. The last four columns show theoretical predictions obtained with NLO QCD, $k_{\rm T}$ factorization, and with the {\sc pythia} and the {\sc sherpa} event generators.}
\label{tab:ratio}
\begin{tabular}{cccccccccc} \hline \hline
~$\Ptg$ bin~ & ~$\la\Ptg\ra$~ & \multicolumn{8}{c}{$\sigma(\gamma+2~b)/\sigma(\gamma+b)$ } \\\cline{3-10}
  (GeV) & (GeV) & Data & $\delta_{\rm stat}$($\%$) & $\delta_{\rm syst}$($\%$) & $\delta_{\rm tot}$($\%$) & ~~~NLO~~~ &~~~$k_{\rm T}$ fact.~~~ & ~~~{\sc pythia}~~~ & ~~~{\sc sherpa}~~~\\\hline

  30 --  40 &   34.5 & ~~1.48$\times 10^{-1}$ &2.3 &+14/$-6$~~ & +14/$-6$~~&1.58$\times 10^{-1}$~~&1.42$\times 10^{-1}$~~&7.25$\times 10^{-2}$~~&8.42$\times 10^{-2}$ \\
  40 --  50 &   44.6 &  ~~1.68$\times 10^{-1}$ & 2.5&+13/$-7$~~ & +13/$-8$~~&2.04$\times 10^{-1}$~~&1.89$\times 10^{-1}$~~&1.18$\times 10^{-1}$&1.20$\times 10^{-1}$\\ 
  50 --  65 &   56.6 & ~~2.36$\times 10^{-1}$ & 2.8 &+12/$-8$~~ & +12/$-8$~~&2.59$\times 10^{-1}$~~&2.34$\times 10^{-1}$~~&1.22 $\times 10^{-1}$&1.63$\times 10^{-1}$\\
  65 --  90 &   75.2 &  ~~2.54$\times 10^{-1}$ & 3.3&+11/$-8$~~ & +12/$-10$~~&3.05$\times 10^{-1}$~~&2.92$\times 10^{-1}$~~&1.30$\times 10^{-1}$&2.20$\times 10^{-1}$\\
  90 -- 200 &   118.3 &  ~~3.14$\times 10^{-1}$ & 3.4& +15/$-14$~~ & +15/-15~~&3.52$\times 10^{-1}$~~&3.67$\times 10^{-1}$~~&1.36$\times 10^{-1}$&2.87$\times 10^{-1}$\\
\hline \hline
\end{tabular}
\end{table*}

The measured cross sections are well described by the NLO QCD calculations 
and the predictions from the $k_T$-factorization approach in the full studied $\Ptg$ region
considering the experimental and theoretical uncertainties.
Both of these predictions show consistent behavior, although the predictions
from the $k_T$-factorization approach suffer from larger uncertainties. {\sc pythia} predicts significantly lower production rates and a more steeply falling $\Ptg$ distribution than observed in data. {\sc sherpa} performs better in describing the normalization at high $\Ptg$, but underestimates production rates compared to that observed in data at low $\Ptg$.

\begin{figure}[htbp]
\includegraphics[width=1.03\linewidth]{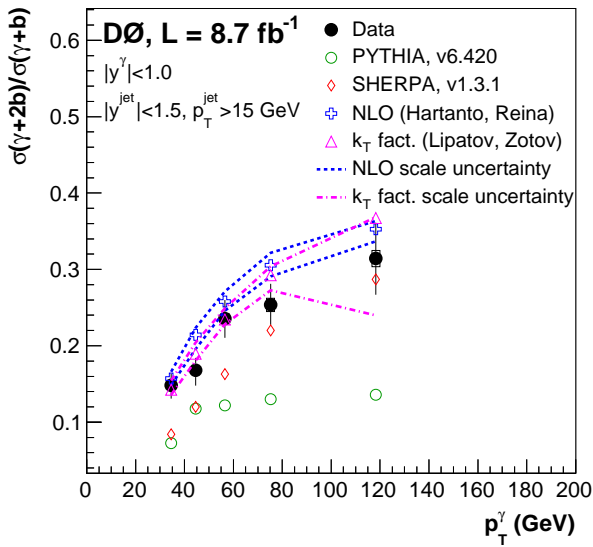}
\caption{ (Color online) The ratio of measured cross sections for $\gamma+2~b$-jet to $\gamma+b$-jet production as a function 
of $\Ptg$ compared to theoretical predictions.
The uncertainties on the data points include both statistical (inner error bar)
and the full uncertainties (full error bar). The measurements are compared to the NLO QCD calculations~\cite{Reina}. The predictions from {\sc sherpa}~\cite{Sherpa}, {\sc pythia}~\cite{PYT} and  $k_{\rm T}$-factorization \cite{Zotov,Zotov2}
  are also shown along with the scale uncertainties on NLO and $k_{\rm T}$-factorization predictions. }

\label{fig:xsect_ratio_cc}
\end{figure}

In addition to measuring the $\gamma+2~b$-jet cross sections, we also obtain results for the
inclusive $\gamma+b$-jet cross section in the same $\Ptg$ bins.
Here we follow the same procedure as described in the previous similar D0 measurement~\cite{gamma_b_d0_2}.
However, as for the $\gamma+2~b$-jet cross section measurement,
we now use the most recent HF tagging algorithm \cite{b-NIM}.
The measured cross sections are shown in Fig.~\ref{fig:xsectLOGplot_gb},
and are compared to various predictions in Fig.~\ref{fig:xsectratio_gb}. 
Data and predictions are also presented in Table~\ref{tab:xsect_gb}.
The values of the obtained $\gamma+b$-jet cross section are consistent 
with our previously published results~\cite{gamma_b_d0_2}. 

We use  $\sigma$($\gamma+2~b$-jet) and $\sigma$($\gamma+b$-jet) cross sections  to calculate their ratio 
in bins of $\Ptg$. Figure~\ref{fig:xsect_ratio_cc} shows
the $\Ptg$ spectrum of the measured ratio.
The systematic uncertainties on the ratio vary within ($11-15$)\%, 
being largest at high $\Ptg$. The major sources of systematic uncertainties 
are attributed to the jet acceptances and the estimation of $b$-jet and $2~b$-jet
fractions obtained from the template fits to the data. Figure~\ref{fig:xsect_ratio_cc} also shows
comparisons with various predictions. The measurements are well described by the calculations done by  NLO
QCD and $k_{\rm T}$-factorization predictions taking into account the experimental and theoretical uncertainties. 
The scale uncertainties on the NLO calculations are typically $\lesssim 15\%$, while they vary uptp $35$\% at high $\Ptg$ for the $k_{\rm T}$-factorization approach. 
The predictions
from {\sc sherpa} describe the shape, but underestimate the ratio for most of the $\Ptg$ bins. The {\sc Pythia} model does not perform well in describing 
the shape and underestimates ratios across all the bins.
Experimental results as well as theoretical predictions for the ratios are presented in Table \ref{tab:ratio}.

In summary, we have presented the first measurement of the differential cross section 
of inclusive production of a photon in
association with two $b$-quark jets as a function of \ptg at the Fermilab Tevatron $p\bar{p}$ Collider. 
The results cover the kinematic range $30<\ptg<200~\GeV$, 
$|y^\gamma|<1.0$, $p_T^\text{jet}>15$ GeV, and $| y^\text{jet}|\lt 1.5$. 
The measured cross sections are in agreement with the NLO QCD calculations and 
predictions from the $k_T$-factorization approach.
We have also measured the ratio of differential $\sigma$($\gamma+2~b$-jet)/$\sigma$($\gamma+b$-jet)
in the same \ptg range. The ratio agrees with the predictions from NLO QCD and $k_{\rm T}$-factorization approach within the theoretical
and experimental uncertainties in the full studied \ptg range.
These results can be used to further tune theory, MC event generators and improve the description of background processes 
in studies of the Higgs boson and searches for new phenomena beyond the Standard Model
at the Tevatron and the LHC 
in final states involving the production of vector bosons in association with two $b$-quark jets.
  
We are grateful to the authors of the theoretical calculations,
H.~B.~Hartanto, L.~Reina, A.~Lipatov and N.~Zotov, 
for providing predictions and for many useful discussions.

%
We thank the staffs at Fermilab and collaborating institutions,
and acknowledge support from the
DOE and NSF (USA);
CEA and CNRS/IN2P3 (France);
MON, Rosatom and RFBR (Russia);
CNPq, FAPERJ, FAPESP and FUNDUNESP (Brazil);
DAE and DST (India);
Colciencias (Colombia);
CONACyT (Mexico);
NRF (Korea);
FOM (The Netherlands);
STFC and the Royal Society (United Kingdom);
MSMT and GACR (Czech Republic);
BMBF and DFG (Germany);
SFI (Ireland);
The Swedish Research Council (Sweden);
and
CAS and CNSF (China).

\bibliography{paper_gbb}

\begin{thebibliography}{99}
\bibitem{Owens} 
 J.~F.~Owens, Rev. Mod. Phys. {\bf 59}, 465 (1987).
\bibitem{Tzvet} 
 T.~Stavreva and J.~F.~Owens, Phys. Rev. D {\bf 79}, 054017 (2009).
\bibitem{gamma_b_d0_2}
V.~M.~Abazov {\sl et al.} (D0 Collaboration), Phys. Lett. B {\bf 714}, 32 (2012).
\bibitem{Reina} 
H.~B.~Hartanto and L.~Reina, Phys. Rev. D {\bf 89}, 074001 (2014).
\bibitem{ttgam}
U.~Baur, A.~Juste, L.~Orr, and D.~Rainwater, Phys. Rev. D {\bf 71}, 054013 (2005). 
\bibitem{gamma_b_d0_1}
V.~M.~Abazov {\sl et al.} (D0 Collaboration), Phys. Rev. Lett. {\bf 102}, 192002 (2009).
\bibitem{gamma_b_d0_3}
V.~M.~Abazov {\sl et al.} (D0 Collaboration), Phys. Lett. B {\bf 719}, 354 (2013).
\bibitem{gamma_b_cdf_1}
T.~Aaltonen {\sl et al.} (CDF Collaboration), Phys. Rev. D {\bf 81}, 052006 (2010).
\bibitem{gamma_b_cdf_2}
T.~Aaltonen {\sl et al.} (CDF Collaboration), Phys. Rev. Lett. {\bf 111}, 042003 (2013).
\bibitem{d0lumi}
T.~Andeen {\sl et al.}, FERMILAB-TM-2365 (2007).

\bibitem{diphoton}
V.~M.~Abazov {\sl et al.} (D0 Collaboration), Nucl. Instrum. Methods Phys. Res. A {\bf 750}, 78 (2014)
\bibitem{hgg_prl}
V.~M.~Abazov {\sl et al.} (D0 Collaboration), Phys. Rev. Lett. {\bf 102}, 231801 (2009).
\bibitem{b-NIM}
V.~M.~Abazov {\sl et al.} (D0 Collaboration), arXiv:1312.7623 (submitted to Nucl. Instrum. Methods Phys. Res. A).

\bibitem{d0det}
V.~M.~Abazov {\sl et al.} (D0 Collaboration), Nucl. Instrum. Methods Phys. Res. A {\bf 565}, 463 (2006);
R.~Angstadt {\sl et al.}, Nucl. Instrum. Methods Phys. Res. A {\bf 622}, 298 (2010);
M.~Abolins {\sl et al.}, Nucl. Instrum. Methods Phys. Res. A {\bf 584}, 75 (2008).
\bibitem{d0_coordinate}
The polar angle $\theta$ and the azimuthal angle $\phi$ are defined
with respect to the positive $z$ axis, which is along the proton beam direction.
Pseudorapidity is defined as $\eta=-\ln[\tan(\theta/2)]$.
Also, $\eta_{\text {det}}$ and $\phi_{\text {det}}$ are the pseudorapidity and the azimuthal angle
measured with respect to the center of the detector.
\bibitem{Geant}
R.~Brun and F.~Carminati, CERN Program Library Long Writeup, W5013, (1993);
we use {\sc geant} version v3.21.
\bibitem{pv}
V.~M.~Abazov {\sl et al.} (D0 Collaboration), Nucl. Instrum. Methods Phys. Res. A  {\bf  620}, 490 (2010).
\bibitem{Sherpa}
T.~Gleisberg {\it et al.}, J. High Energy Phys.   02, 007 (2009).
We use {\sc sherpa} version v1.3.1.

\bibitem{PYT}
T.~Sj\"ostrand, S.~Mrenna, and P.~Z.~Skands, J. High Energy Phys.  05, 026 (2006).
We use {\sc pythia} version v6.420 with tune A.

\bibitem{Run2Cone}
G.~C.~Blazey {\sl et al.}, arXiv:hep-ex/0005012 (2000).
\bibitem{PhotonInc}
V.~M.~Abazov {\sl et al.} (D0 Collaboration),
Phys. Lett. B {\bf 639}, 151 (2006). 
\bibitem{Frag}
T.~Binoth {\it et al.}, Eur. Phys. J. C {\bf 4}, 7 (2002).
\bibitem{DMJL_wb}
V.~M.~Abazov {\sl et al.} (D0 Collaboration), Phys. Lett. B {\bf 718}, 1314 (2013).
\bibitem{particle}
C.~Buttar {\sl et al.}, arXiv:0803.0678 [hep-ph], section 9.

\bibitem{JES} 
V.~M.~Abazov {\sl et al.} (D0 Collaboration), arXiv:1312.6873 (accepted by  Nucl. Instrum. Methods).
\bibitem{TW} 
G.~D.~Lafferty and T.~R.~Wyatt, Nucl. Instrum. Methods Phys. Res. A  {\bf  355}, 541 (1995).
\bibitem{CTEQ}
W.~K.~Tung {\sl et al.}, J. High Energy Phys. {\bf 02}, 052 (2007).


\bibitem{Zotov}
A.~V.~Lipatov and N.~P.~Zotov, 
J. Phys. G {\bf 34}, 219 (2007);
S.~P.~Baranov, A.~V.~Lipatov, and N.~P.~Zotov, Eur. Phys. J. C {\bf 56}, 371  (2008).
\bibitem{Zotov2}
A.~V.~Lipatov and N.~P.~Zotov, paper in preparation.
\bibitem{UPD}
M.~A.~Kimber, A.~D.~Martin, and M.~G.~Ryskin, Phys. Rev. D {\bf 63}, 114027 (2001).
\bibitem{SHERPA_gam}
A.~D.~Martin, W.~J.~Stirling, R.~S.~Thorne, and G.~Watt, Eur. Phys. J. C {\bf 63}, 189 (2009).
\bibitem{Sherpa_scale}
We  choose the following ME-PS matching parameters: the energy scale 
$Q_0 = 15$ GeV and the spatial scale $D = 0.4$, where $D$ is taken to
be of the radius of the photon isolation cone.



\end{thebibliography}


\begin{thebibliography}{99}
\end{thebibliography}
\bibliographystyle{apsrev}

\end{document}